\documentclass[aps,showpacs,twocolumn]{revtex4}

\usepackage{graphicx}
\usepackage{bm}
\usepackage{amsmath}

\newcommand{\be}{\begin{eqnarray}}
\newcommand{\ee}{\end{eqnarray}}

 \newcommand{\gsim}{\mathrel{\hbox{\rlap{\lower.55ex \hbox {$\sim$}}
                   \kern-.3em \raise.4ex \hbox{$>$}}}}
\newcommand{\lsim}{\mathrel{\hbox{\rlap{\lower.55ex \hbox {$\sim$}}
                   \kern-.3em \raise.4ex \hbox{$<$}}}}

\newcommand{\ba}{\begin{eqnarray}}
\newcommand{\ea}{\end{eqnarray}}

\usepackage{amssymb}
\newcommand{\rhoc}{\rho_{\mathrm{c}}}
\newcommand{\rhop}{\rho_{\mathrm{p}}}
\newcommand{\rhopcn}[1]{\rho_{\mathrm{p},#1}}

\newcommand{\rhopn}[1]{\rho_{\mathrm{p},#1}}

\newcommand{\rhopzero}{\rho_{\mathrm{p},0}}
\newcommand{\drhoc}{\delta \rho_{\mathrm{c}}}
\newcommand{\drhop}{\delta \rho_{\mathrm{p}}}
\newcommand{\T}[1]{T_{#1}}
\newcommand{\Tzero}[1]{T_{#1}^{0}}
\newcommand{\dT}[1]{\delta T_{#1}}
\newcommand{\e}[1]{\epsilon_{#1}}

\newcommand{\ave}[1]{< #1 >}

\begin{document}


\title{Jet Tomography of Harmonic Fluctuations \\
in the Initial Condition of Heavy Ion Collisions}
\author{Xilin Zhang$^{1}$ and Jinfeng Liao$^{1,2}$}
\address{$^1$ Physics Department and Center for Exploration of Energy and Matter,
Indiana University, 2401 N Milo B. Sampson Lane, Bloomington, IN 47408, USA.\\
$^2$ RIKEN BNL Research Center, Bldg. 510A, Brookhaven National Laboratory, Upton, NY 11973, USA.}
\date{\today}

\begin{abstract}
In this paper we study the jet response (particularly azimuthal anisotropy) as a hard probe of the harmonic fluctuations in the initial condition of central heavy ion collisions. By implementing the fluctuations via cumulant expansion for various harmonics quantified by $\epsilon_n$ and using the geometric model for jet energy loss, we compute the response $\chi^h_n=v_n/\epsilon_n$. Combining these results with the known hydrodynamic response of the bulk matter expansion in the literature, we show that the hard-soft azimuthal correlation arising from their respective responses to the common geometric fluctuations reveals a robust and narrow near-side peak that may provide the dominant contribution to the ``hard-ridge'' observed in experimental data.
\end{abstract}
\maketitle

\section{Introduction}

The structure and properties of QCD matter under extremely hot and/or dense  conditions are of fundamental interest and provide unique environments for
studying the strongest force of Nature. The hot deconfined QCD matter, the so-called quark-gluon plasma (QGP), was part of the history for cosmic evolution after the Big Bang and has now been created via relativistic heavy ion collisions (the ``Little Bang'') and explored in laboratory experiments at the Relativistic Heavy Ion Collider (RHIC) and the Large Hadron Collider (LHC).
In such collisions, highly energetic jets born from initial hard  scattering
provide natural ``tomography'' of the created hot QCD matter. Jet quenching due to energy loss along
the jet path through the medium encodes essential information about
the dynamics of jet-medium interaction and the medium properties as well,
which shall be inferrable from experimental observables such as the  high-$p_t$ hadron suppression and azimuthal anisotropy, as well as the di-hadron correlations (for reviews see e.g. \cite{Gyulassy:2003mc}).

While the jet quenching has been experimentally established as a very robust
 phenomenon at RHIC and now at LHC, the microscopic mechanism of jet energy
 loss is not yet fully understood. A conventional observable for quantifying
 jet quenching is the nuclear modification factor $R_{AA}$ which compares
 the particle production in the AA collision to the naive expectation
 from simply scaling up single NN cross section by the binary NN collision
 number. A measured $R_{AA}$ significantly smaller than unity for the high-$p_t$
 hadrons implies strong in-medium energy loss for the jets: indeed for central
 collisions at both RHIC and LHC we've seen $R_{AA}\le 0.2$. The $R_{AA}$
 provides direct information on the {\em average} opaqueness of the created
 hot medium, and it is customary in various jet quenching models to use $R_{AA}$
 in the most central collisions to normalize their respective parameters
 for the average jet-medium interaction strength. More sensitive are the
 geometric features of jet quenching observables that are particularly useful
 in discriminating different models of energy loss. These include the
 $A$-dependence (when changing colliding systems), the $b$-dependence
 (when changing the collision impact parameter or centrality class), and
 the $\phi$-dependence (when changing the probe jet's azimuthal orientation with respect to
 the reaction plane). For any given model with its parameters fixed in
 the most central collisions, the above geometric dependence and the
 correlations among different observables then provide crucial tests
 of the model \cite{Gyulassy:2000gk,Shuryak:2001me,Liao:2008dk,Jia:2010ee,Jia:2011pi,Betz:2011tu,Liao:2011kr,Adler:2006bw,Adare:2010sp}.

Let's elaborate a bit on the $\phi$-dependence of $R_{AA}$. In non-central collisions, the medium ``thickness'' as seen by a penetrating jet depends on the azimuthal angle $\phi$ of the jet with respect to the reaction plane, therefore leading to the reaction-plane dependence of high-$p_t$ hadron suppression i.e. $R_{AA}(\phi)$ \cite{Gyulassy:2000gk}. The dominant anisotropy in $R_{AA}(\phi)$ (for non-central collisions) can be attributed to the second harmonic term $\cos(2\phi-2\Psi_{R.P.})$ with its coefficient being the elliptic ``flow'' parameter for high-$p_t$ hadrons, $V_2^{hard}$ which
is a non-collective component though \cite{Liao:2009ni}. Despite the success of many models in describing $R_{AA}$ and its centrality dependence, most models significantly under-predicted the $V_2^{hard}$ and failed the test by geometric data \cite{Shuryak:2001me}\cite{Adler:2006bw}. The lack of a simultaneous description  for $R_{AA}$ and $V_2^{hard}$ in a single model was not resolved till a new insight suggested in \cite{Liao:2008dk}. Motivated by the ``magnetic scenario'' for sQGP \cite{Liao:2006ry}, the authors of \cite{Liao:2008dk} pointed out that the energy loss of a jet may not simply scale with the local medium density as most models have assumed, but actually has nontrivial dependence on matter density (or temperature). It was particularly shown that the geometric data $R_{AA}$ and $V_2^{hard}$ versus centrality can be successfully described together by including a jet quenching component with strong enhancement in the near-$T_c$ matter by a factor $3\sim 5$ compared with higher-$T$ QGP. Such an enhancement of jet-medium interaction may originate from  non-perturbative structures created by the (color-)electric jet passing a plasma of (color-)magnetic monopoles that dominate the near-$T_c$ matter \cite{Liao:2006ry}\cite{Liao:2008vj}. A natural prediction of this scenario is  that the effective jet-medium interaction would be rapidly reduced when going from RHIC to LHC energies, or the hot matter created at LHC will be more ``transparent'' (apart from the trivial density factor) to a penetrating jet at LHC as compared with that at RHIC. Interestingly, recent quantitative computations and comparison with LHC data at $\sqrt{s}=2.76\rm TeV$ \cite{Horowitz:2011gd}\cite{Betz:2012qq}\cite{Liao:2011kr} indeed  suggest that 1) fixing the jet-medium interaction at RHIC and applying it directly to LHC would lead to ``over-quenching'' as compared with data; 2) reducing the jet-medium interaction by about a factor of 2 would allow a good description of the LHC data --- such reduction is remarkably rapid provided only a $30\%$ increase in temperature from RHIC ($\sqrt{s}=0.2 \rm TeV$) to current LHC ($\sqrt{s}=2.76\rm TeV$)! The future LHC heavy ion data at $\sqrt{s}=5.5\rm TeV$ will be essential in establishing how much and how fast the jet-medium interaction decreases with temperature.

In this paper we explore a new and interesting geometric aspect of jet quenching: the hard probe of the geometric fluctuations (in terms of various harmonics in azimuthal angle) in the initial condition of heavy ion collisions. It has recently been shown in measurements \cite{Lacey:2011av,Sorensen:2011fb,GrosseOetringhaus:2011kv,Jia:2011hfa,Li:2011mp} and demonstrated in various  modelings \cite{Alver:2010gr,Alver:2010dn,Luzum,Luzum:2011jpg,Sorensen:2008bf,Teaney:2010vd,Qiu:2011iv,Staig:2010pn,Takahashi:2009na,Schenke:2010rr,Xu:2010du,Qin:2010pf,Ma:2010dv} that there are very strong fluctuations in the initial matter profile from event to event. Such fluctuations contain various higher order harmonics in azimuthal angle $\sim \cos(n\phi)$ rather than the naive expectation of $2nd$-harmonic dominance in averaged geometry. In the collective expansion of the bulk matter, these fluctuations lead to observed harmonic flows up to about $n=6$ and also explain the soft di-hadron azimuthal correlations (the ``soft-ridge''). In particular, it was shown in \cite{Alver:2010dn} that there is primarily a linear response of the low-$p_t$ hadrons' harmonic flows to the initial harmonic fluctuations. A natural question to ask is, therefore, how a penetrating jet responds to the strongly fluctuating matter density from event to event. There have been some discussions in the literature \cite{Jia:2010ee}\cite{Betz:2011tu}\cite{Rodriguez:2010di} and also data from both RHIC\cite{Adare:2011tg}\cite{Sorensen:2011fb} and LHC\cite{Chatrchyan:2012wg}\cite{Aad:2012bu}, but a systematic and clear picture has not been achieved. Intuitively one would expect the event-by-event azimuthal distribution of high-$p_t$  hadrons shall also reflect such harmonic fluctuations in the initial condition and it is our purpose to systematically quantify such response. In addition, both the quenching of hard jet and the expansion of soft matter are commonly correlated to the same underlying particular matter profile in a given event and such hard-soft correlation will survive the event average. We will examine how such correlation contributes to the di-hadron azimuthal correlation with a hard trigger and a softer associate hadron, in particular a possible explanation of the ``hard-ridge'' \cite{STAR_ridge} for which the origin has so far not been fully understood \cite{Shuryak:2007fu}\cite{Majumder:2006wi}\cite{Dumitru:2008wn}.
To highlight the role of pure fluctuations in leading to azimuthal anisotropy, we limit our study in the present paper only for the most central collisions, and to be specific we do calculations for RHIC $\sqrt{s}=200\rm GeV$ collisions. Unlike in non-central collisions with strong anisotropy already in the average geometry and dominated by the $2nd$ harmonics, in the perfectly central collisions ($b=0$) the average background geometry is isotropic and the anisotropy from various harmonic fluctuations will be best manifested.

The paper is organized as follows. In section II we will describe our model setup for the present study. The simulation results for jet response to the harmonic fluctuations in central collisions, characterized by $v_n/\epsilon_n$ for $n=1,2,...,6$, will be presented in section III. Such results will be used  in section IV to compute the hard-soft di-hadron correlations due to harmonic fluctuations where we will find a robust and narrow near-side peak in the azimuthal angle $\Delta\phi$ dependence as a possible explanation of the so-called ``hard-ridge''. The summary and discussions will be given in section V.

\section{Model Setup}

In this section we introduce our model setup. First we will show how we implement various harmonic fluctuations on top of the isotropic fireball in a perfectly central collision by using a modified cumulant expansion based on that of \cite{Teaney:2010vd}. In particular we will discuss the fluctuations in both the participant density and the binary collision density profiles and their mutual relations. Second we will briefly discuss the geometric model we use for calculating the jet energy loss and its azimuthal angle dependence for a given matter distribution.

\subsection{Parametrization of Harmonic Fluctuations in Central Collisions}

In each central collision event, the participant density ($\rhop$) and the binary collision density ($\rhoc$) are related to the thickness functions of the two colliding nuclei ($\T{a,b}$):
$\rhoc(\vec{r}) = \sigma \T{a} (\vec{r}) \, \T{b}(\vec{r})$,
$\rhop(\vec{r})= \T{a}(\vec{r}) \left[1-P_{b}(\vec{r})\right]+(a\leftrightarrow b)$.
Here $\sigma (=42 \ \mathrm{mb})$ is the N-N inelastic scattering cross section at $\sqrt{s}=200\rm GeV$,
 $A$ is the number of nucleons in each nucleus, and $P_{b}= \left(1-\sigma \T{b}/A \right)^{A}$. It should be emphasized that in general one shall use $\T{a} (\vec{r}-\vec{b}/2)$ and $\T{b} (\vec{r}+\vec{b}/2)$, and the above is only to be used for the perfectly central collisions with $b=0$. Suppose event by event, $\T{a,b}$ fluctuate around an isotropic background $\T{0}$, i.e. $\T{a,b}=\Tzero{a,b}+\dT{a,b}$. As the result, by using the linear approximation,
 the fluctuations of $\rhoc$ and $\rhop$ are directly related, as shown by
\be
\drhoc &=& \sigma \Tzero{} \, (\dT{a}+\dT{b}) \ , \label{eqn:deltarhoc} \\
\drhop &=& \left[1-\left(1-\frac{\sigma \Tzero{}}{A}\right)^{A} +\sigma \Tzero{} \left(1-\frac{\sigma \Tzero{}}{A}\right)^{A-1} \right] \notag \\
         &{}& \times (\dT{a}+\dT{b}) \ .  \label{eqn:deltarhop}
\ee
$T^0$ is obtained by the (optical) Glauber model. To parameterize the fluctuations, we focus on $\drhop$ and later
use the above equations to compute the corresponding  $\drhoc$.

We will use a modified cumulant expansion based on the methods in \cite{Teaney:2010vd,Luzum:2011jpg}. Generally, we can make a Fourier analysis for any profile
\begin{eqnarray}\label{Eq_general_profile}
\rhop(\vec{r})= \rhopzero(r)+\sum_{n\geqslant1} 2\rhopcn{n}(r)\cos\left[{n(\phi-\Psi_n)} \right]
\end{eqnarray}
The difficulty is to  parameterize the function form for the coefficients $\rhopcn{n}(r)$. The small-$k$ expansion method \cite{Teaney:2010vd,Luzum:2011jpg} starts with the transformation (for 2-D transverse plane) $\rhop(\vec{k})\equiv \int d\vec{r} \rhop(\vec{r}) \exp{(i\vec{k}\vec{r})}$, and examines instead the Fourier series $\rhopzero(k)+\sum_{n\geqslant1}2\rhopcn{n}(k)\cos\left[{n(\phi_{k}-\Psi_n^k)}\right]$. One can then make a small-$k$ expansion of the coefficients $\rhopcn{n}(k)$ like e.g. $\rhopcn{n}(k)=\sum_{m\geqslant0}\rhopcn{m,n}(ik)^{m}/m!$. The coefficient $\rhopzero(k)$ is irrelevant to the azimuthal anisotropy and  will not be addressed here. It is easy to see that \cite{Luzum:2011jpg} for the $n-th$ harmonic ($n\geqslant 2$), the leading term in $\rhopcn{n}(k)$'s $k$ expansion is $\rhopcn{n,n}$($=1/2^{n} \ave{r^{n} \cos{\left[{n(\phi-\Psi_{n})}\right]} }$, $\ave{\cdot \cdot}$ means averaging over {\em normalized} $\rhop{(\vec{r})}$). The case for $n=1$ deserves special treatment: it starts with $\rhopcn{3,1}$ ($=3/8 \ave{r^{3} \cos{(\phi-\Psi_{1})} }$) as one can set $\rhopcn{1,1}=0$ by choosing the ``right origin'' of coordinates. Practically one then has to truncate the series, e.g. keeping only leading terms (assuming the relevant $k$ is small). Due to the truncation, upon transforming back to the $\vec r$-space one needs to regulate the large $r$ part. Different from previous approaches, we will simply use a Gaussian factor with a length parameter $\Sigma$ that roughly reflects the scale of fluctuations. In the end, we have the following parameterizations for various orders of harmonics:
\begin{eqnarray} \label{Eq_cumulant}
\rhopn{1}(r)&=& - \frac{\e{1}}{2}\frac{\ave{r^{3}}}{\Sigma^{3}} \left[\left(\frac{r}{\Sigma}\right)^{3}-2\left(\frac{r}{\Sigma}\right)\right] \frac{1}{\pi \Sigma^{2}} e^{-\frac{r^{2}}{\Sigma^{2}}}  \ , \quad \\
\rhopn{n}(r)&=&
-\frac{\e{n}}{n!}  \frac{\ave{r^{n}}}{\Sigma^{n}}  \left(\frac{r}{\Sigma}\right)^{n} \frac{1}{\pi \Sigma^{2}} e^{-\frac{r^{2}}{\Sigma^{2}}} \ ,   \ n\geqslant2 \ .
\end{eqnarray}
Here $\e{1}\equiv-\ave{r^{3}\cos{(\phi-\Psi_{1})}}/\ave{r^{3}}$, $\e{n\geqslant2}\equiv-\ave{r^{n}\cos{n(\phi-\Psi_{n})}}/\ave{r^{n}}$, $\ave{r^{n}}^{1/n}=3.88, 4.19, 4.44, 4.63, 4.80, 4.95\, \rm fm$ for $n=1,2,3,4,5,6$. The root-mean-radius-square $\sqrt{\ave{r^{2}}}=\, 4.19\, \rm fm$ is defined as $\sigma$ and will be used as a ``ruler'' for $\Sigma$.
Therefore by specifying the values of $\epsilon_n$ (as well as $\Psi_n$) for an event, one then fixes the  $\rhopn{n}(r)$ above and obtains a particular density profile in Eq.(\ref{Eq_general_profile}) with harmonic fluctuations.

 One technical complication is that for such parametrization at large radius, the total density can become negative. Hence certain regularization scheme has to be implemented as discussed in details by \cite{Teaney:2010vd}. We simply set the density beyond that critical radius as zero. For the condition of our interest, i.e. $\epsilon \approx 0.1$, this critical radius is around $7$ fm and our regularization is physical. The regularization will require a re-calibration of the eccentricities: all the actual $\epsilon_n$ values need to be evaluated directly from the generated and regularized profile which would generally differ from the input $\epsilon_n$ parameters in Eq.(\ref{Eq_cumulant}), see e.g. \cite{Teaney:2010vd}\cite{Luzum}.

\subsection{The Jet Energy Loss}

In this study, we use the geometric model approach for computing the jet energy loss. Such models reflect the generic geometric features (e.g. the path-length dependence) that are most crucial for describing geometric data \cite{Shuryak:2001me,Liao:2008dk,Jia:2010ee,Jia:2011pi,Betz:2011tu,Liao:2011kr}. We assume that the final energy $E_f$ of a jet with initial energy $E_i$ after traveling an in-medium path $\vec{P}$ can be parameterized as $E_f = E_i \times f_{\vec P}$ with the suppression factor $f_{\vec P}$ given by:
\begin{equation} \label{Eq_fP}
f_{\vec P} = exp\left\{ - \int_{\vec P}\, \kappa[s(l)]\, s(l)\, l^m dl  \right\}
\end{equation}
In the above the $s(l)$ is the entropy density of local matter at a given point on the jet path, while the $\kappa(s)$ is the local jet quenching strength which as a property of matter should in principle depend on the local density $s(l)$. There can be different choices of the parameter $m$ for path-length dependence (e.g. LPM-motivated quadratic or AdS/CFT-motivated cubic) and of the jet-medium interaction $\kappa(s)$. In the present study, we use the near-$T_c$ enhancement model as in \cite{Liao:2008dk}\cite{Liao:2011kr}, assuming $m=1$ and introducing a strong jet quenching component in the vicinity of $T_c$ (with density $s_c$ and span of $s_w$) via
\begin{eqnarray} \label{Eq_kappa}
\kappa(s)=\kappa [1+ \xi\, exp(-(s-s_c)^2/s_w^2)]
\end{eqnarray}
with $\xi=6$, $s_c=7/fm^3$, and $s_w=2/fm^3$. (see \cite{Liao:2008dk} for  details.)
As aforementioned, the parameter $\kappa$ will be fixed by $R_{AA}\approx 0.18$ in the $0-5\%$ collisions at RHIC $\sqrt{s}=200\rm GeV$. 

It shall be emphasized that the jet path $\vec{P}$ is determined by the initial jet spot coordinates  on the transverse plane as well as the the azimuthal angle $\phi$ for the transverse orientation of its propagation.
After averaging over all jet paths (including all possible start points properly weighed by binary collision density and all equally possible orientations) one may then obtain the $R_{AA}$:
\begin{equation}
R_{AA} = <\, (f_{\vec P})^{n-2}  \,>_{\vec P}
\end{equation}
where the exponent $n$ comes from measured reference p-p spectrum (see e.g. \cite{Adler:2006bw} for a detailed account). The value of $n$ depends on collision energy and throughout this paper we focus on the RHIC $\sqrt{s}=200\rm GeV$ collisions with $n\approx 8.1$. Note that in this model as in many other geometric models, assuming the fractional energy loss will lead to the $R_{AA}$ independent of $p_t$ which may be justified by the approximate ``flatness'' seen in the data at RHIC energy. The results from such modeling apply only to the high-$p_t$ region, e.g. $p_t>6\, \rm GeV$ at RHIC.  

The azimuthal angle $\phi$-dependence can be studied by averaging over jet paths with a particular azimuthal orientation, i.e. $R_{AA}(\phi) = <\, (f_{\vec P})^{n-2}  \,>_{{\vec P}(\phi)}$ from which various harmonic components are derivable by  Fourier decomposition.
In this study for each given geometry, we compute $R_{AA}(\phi)$ as a function of azimuthal angle $\phi$ by integrating over all initial jet spots on the transverse plane weighed by the binary collision density at each spot for a given value of $\phi$.

\section{Jet Response to Harmonic Fluctuations}

\begin{figure*}[t]
\centering
\includegraphics[width=9.3cm, angle=-90]{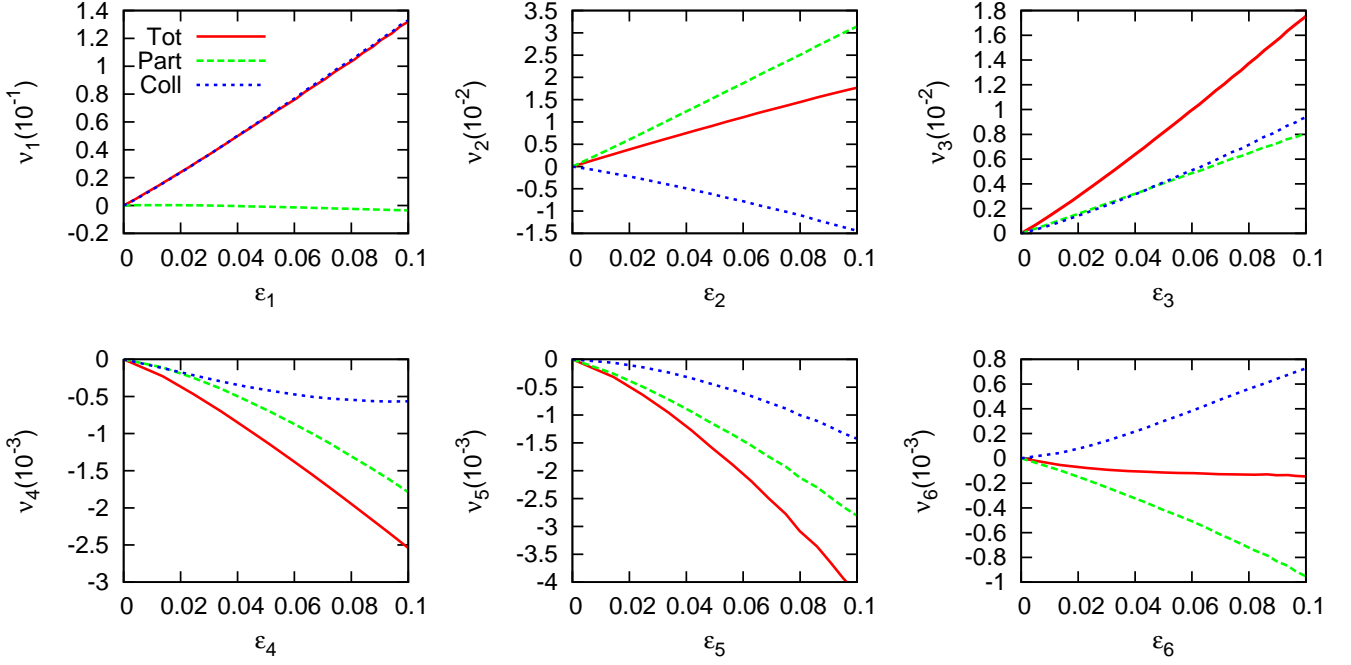}
  \caption{(color online) The calculated $\nu_{n}$ vs. $\e{n}$ for various harmonic fluctuations $n=1,2,3,4,5,6$ with parameter $\Sigma$=$\sigma$.  Three different responses are shown: the case with only $\rhop$ fluctuation (``Part'', green dashed curves), the case with only $\rhoc$ fluctuation (``Coll'', blue dotted curves), and the physical case with both (``Tot'', red solid curves) (see text for details).}
  \label{fig:nu_vs_e}
\end{figure*}

Here we report our results for the jet response to harmonic fluctuations: see Fig.~\ref{fig:nu_vs_e}. The results are obtained with $\Sigma=\sigma=4.19$ fm, which indicates the scale of the fluctuation is roughly the same as that of the isotropic background. In our approach, the fluctuations $\epsilon_n$ are assumed to be small so that the jet response to density-fluctuations is approximately linear. We will first demonstrate that indeed for $n-th$ harmonics alone with amplitude $\e{n}$, the jet response  to a good approximation depends on $\e{n}$ linearly (hence we can define $\chi^h_n=v_n/\epsilon_n$).  The solid curves in each panel clearly show that  for $n=1,2,3$, the relations between $v_{n}$ and $\e{n}$ are linear, while for $n=4,5,6$ only very mild  nonlinearity starts to develop near $\epsilon_n=0$ and the response stays quite linear for the physically most relevant region $0.05< \epsilon_n < 0.1$. For each harmonic fluctuation, we have checked that the initial axis angle  $\Psi_{n}$ of fluctuation coincides with the final axis angle $\Psi_{n}^{J}$ of the jet distribution anisotropy.  

One important issue is to understand the different influences on the jet response due to the  fluctuation in the participant density (which mostly concerns the bulk matter that quenches the jets) and that in the binary collision  density (which concerns the profile of initial jet spots). The difference could be tricky and counter-intuitive (see e.g. some discussions in \cite{Rodriguez:2010di}). Here we clearly demonstrate the difference by showing three different responses in Fig.~\ref{fig:nu_vs_e}: the case with only $\rhop$ fluctuation (``Part'', green dashed curves), the case with only $\rhoc$ fluctuation (``Coll'', blue dotted curves), and the physical case with both (``Tot'', red solid curves). We've checked that the total response agrees well with the sum of the two individual contributions which provides additional evidence for the linear nature of the response. As one can see, the contributions to the jet response due to  fluctuation in participant density and that in collision density differ significantly in the absolute magnitude and can even have opposite signs (in the case of $n=1,2,6$) thus canceling each other to certain extent. It is therefore clear that for event-by-event studies of jet response, both fluctuations have to be fully taken into account.

\begin{figure}
  \includegraphics[width=8cm]{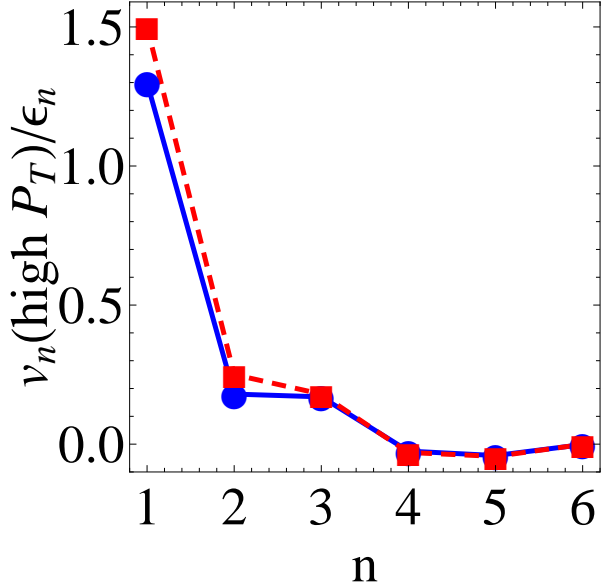}
  \caption{(color online) The jet response $\chi^h_n=v_n/\epsilon_n$ to various orders of harmonic fluctuations in the initial condition, with the blue solid curve for parameter $\Sigma=\sigma$ and the red dashed curve for parameter $\Sigma=1.25\sigma$. (see text for details).}
  \label{fig_response}
\end{figure}

   Finally in the Tab.~\ref{tab:trespsum}, we show the jet response coefficients $\chi^h_n$ extracted from the physically most relevant  region $0.05< \epsilon_n < 0.1$ where the linear dependence is a very well approximation for all harmonics. The results as a response spectrum $\chi^h_n\,$  vs. $\, n$ are also plotted in Fig.~\ref{fig_response}. The response spectrum shows a typical decrease from low to high harmonics, and somewhat surprisingly (as compared with the soft response) there is a very strong response in the first harmonics. We also see the responses are
   relatively insensitivity to the parameter $\Sigma$ and therefore the quantified response spectrum shape in $n$ is robust. While it would be interesting to directly compare with data, currently the RHIC data \cite{Adare:2011tg}\cite{Sorensen:2011fb} for higher harmonics are only available for the low to intermediate $p_t$ region so such comparison with our results would not be appropriate. We point out though the previous comparison between the results from the same model with $v_2$ data at high $p_t$ was rather successful, see e.g. \cite{Liao:2008dk}\cite{Liao:2011kr}.

\begin{table} [!h]
 \centering
   \begin{tabular}{|c|c|c|c|c|c|c|} \hline
 Total Response & $\chi^h_1$ & $\chi^h_2$ & $\chi^h_3$
 & $\chi^h_4$ & $\chi^h_5$  & $\chi^h_6$ \\ \hline
 $\Sigma=\sigma$ & 1.3 & 0.18 & 0.17 & -2.5e-2 &-4.1e-2 &-1.5e-3   \\ \hline
 $\Sigma=1.25\sigma$ & 1.5 & 0.25 & 0.18 & -3.0e-2 & -4.4e-2 &-8.8e-4 \\ \hline
   \end{tabular}
   \caption{Total responses to various orders of harmonic fluctuations in the initial condition using different values for the parameter $\Sigma$ (see text for details).} \label{tab:trespsum}
\end{table}

We end this section by discussing certain check we've done.  One issue is about possible ``mixing'' or ``interfering''  effect among different harmonic fluctuations when they all co-exist with varied respective main-axis for each. This should be clarified as in reality each collision event would automatically come with all these harmonic fluctuations. We've done the following test: we include on top of the isotropic background all the $n=1,2,3,4,5,6$ harmonic fluctuations simultaneously with randomly assigned $\epsilon_n$ (in the linear regime though) and randomly chosen axis $\Psi_n$ (different for each), and then calculate the jet quenching result $R_{AA}(\phi)$. We've found that
1) the main-axis for each harmonics $\Psi_n^{J}$ as determined by maximizing the corresponding $\cos(\phi-\Psi_n^{J})$ component in final state $R_{AA}(\phi)$ agrees with the initial $\Psi_n$ from fluctuation with less than $1\%$ difference; 2) the response $v_n/\epsilon_n$ also agrees with the values we extracted above by treating each harmonics separately. Another issue that we've checked is how such response would change with different jet energy loss models. We've also done the calculation for two other models: the model with path-length-square dependence and constant jet-medium interaction, i.e. $\kappa(s)=\kappa$ and $m=1$ in Eq.(\ref{Eq_fP}); and the model with path-length-cubic dependence and constant jet-medium interaction, i.e. $\kappa(s)=\kappa$ and $m=2$ in Eq.(\ref{Eq_fP}). The response coefficients
$\chi^h_n$ from different models are not drastically different, show similar decreasing trend with increasing $n$, and may become distinguishable when  realistic and accurate data comparison can be done.

\section{The ``Hard-Ridge'' in Di-Hadron Correlations}

In this section, we study the correlation between the hard and soft sector due to their respective  correlations to the common initial condition with fluctuations. It would be interesting to see what major features of the hard-soft correlation structures can arise from the jet response to harmonic fluctuations in the initial condition. 

Suppose for an arbitrary central collision event, we have the initial azimuthal distribution of matter density fluctuations characterized by a series of harmonics with varied (and uncorrelated) magnitude $\epsilon_n$ and axis orientation $\Psi_n$. According to our results on the jet response, we would expect the final state  high-$p_t$ hadron distribution to be of the form
\begin{eqnarray} \label{Eq_hard_dist}
\frac{dN^{hard}}{dy d\phi} \sim\, 1 + \sum_{n=1,2,3,...} 2 v^{h}_n \cos[n(\phi-\Psi_n)]
\end{eqnarray}
The bulk matter, on the other hand, will generate the harmonic flows during hydrodynamic expansion and lead to the final state low $p_t$ hadron distribution of the form
\begin{eqnarray} \label{Eq_soft_dist}
\frac{dN^{soft}}{dy d\phi} \sim\, 1 + \sum_{n=1,2,3,...} 2 v^{s}_n \cos[n(\phi-\Psi_n)]
\end{eqnarray}
Note that for a given event both the soft and the hard responses to each order of the harmonic fluctuations are commonly aligned to the same corresponding angle $\Psi_n$ from the initial condition --- such a correlation between the hard and soft responses will survive the average over many events.

We can therefore define a hard-soft correlation function
\begin{eqnarray}
&& C[\Delta \phi] \equiv  \nonumber \\ && \frac{\int \frac{d\phi_1}{2\pi} \frac{d\phi_2}{2\pi}  2\pi\delta(\phi_2-\phi_1-\Delta\phi) <\frac{dN^{hard}}{dy d\phi_1}\, \frac{dN^{soft}}{dy d\phi_2}>}{<\int \frac{d\phi_1}{2\pi} \frac{dN^{hard}}{dy d\phi_1}>\, <\int \frac{d\phi_2}{2\pi}\frac{dN^{soft}}{dy d\phi_2}>} \,  -1 \qquad
\end{eqnarray}
where the $<\,>$ means averaging over events. Since the initial harmonic fluctuations have their orientation angles $\Psi_n$ vary randomly from event to event and uncorrelated among each other (at least so in central collisions), we obtain from Eq.(\ref{Eq_hard_dist})(\ref{Eq_soft_dist}) the following
\begin{eqnarray}
C[\Delta \phi] = \sum_{n=1,2,3,...} 2\, < v^{h}_n v^{s}_n>\, \cos(n\Delta\phi)
\end{eqnarray}
 In the linear response approximation we can use $v^{h,s}_n=\chi^{h,s} \epsilon_n$ to further get
\begin{eqnarray} \label{Eq_corr_chi}
C[\Delta \phi] = \sum_{n=1,2,3,...} 2\, \chi^{h}\,\chi^{s}\, < (\epsilon_n)^2>\, \cos(n\Delta\phi)
\end{eqnarray}
We therefore see that the common correlations of the hard and soft responses with various harmonic fluctuations indeed lead to the hard-soft correlation that survives the event average.

\begin{figure}
  \includegraphics[width=8cm]{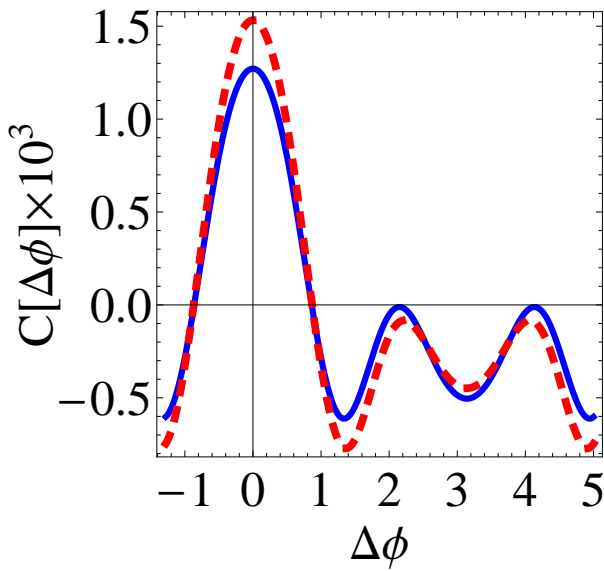}
  \caption{(color online) Hard-soft correlation $C[\Delta\phi]$ from responses to common geometric fluctuations in the initial condition, with the blue solid curve using jet response results from $\Sigma=\sigma$ case  while the red dashed curve from $\Sigma=1.25\sigma$ case (see text for details).}
  \label{Fig_ridge}
\end{figure}

It it tempting to quantitatively examine the features of such hard-soft azimuthal correlation. To do that, we compute $C[\Delta\phi]$ by including the $n=1,2,3,4,5$ harmonics in Eq.(\ref{Eq_corr_chi}). For the initial fluctuations we use results from Monte Carlo simulations of initial conditions for most central collisions, which are $\sqrt{\epsilon_{n=1,2,3,4,5}^2}\approx 0.037,0.068,0.076,0.084,0.091$ respectively (see e.g. \cite{Alver:2010dn}\cite{Luzum:2011jpg}\cite{Teaney:2010vd}\cite{Qiu:2011iv}). For the soft response coefficients we take the results from hydrodynamic modeling, which are $\chi^s_{n=1,2,3,4,5}\approx 0.15,0.26,0.21,0.14,0.086$ respectively (see e.g. \cite{Alver:2010dn}\cite{Luzum:2011jpg}\cite{Teaney:2010vd}\cite{Qiu:2011iv}).
The hard response coefficients are taken from our results: since such coefficients depend on the parameter $\Sigma$ we will show results for both values of $\Sigma$ as studied in the previous section. The so-obtained correlation function $C[\Delta\phi]$ is plotted in Fig.3: the blue solid curve is obtained using $\chi^h_n$ results from $\Sigma=\sigma$ case  while the red dashed curve  is from $\Sigma=1.25\sigma$ case. A few remarks are in order:\\
1) there is a robust and narrow near-side peak around $\Delta\phi=0$. Actually for both cases we've fitted very well the peaks by a Gaussian function with width about $0.8$ radian which is very close to the experimental data. This main feature, we believe, could provide a natural explanation of the ``hard-ridge'' structure as arising from superposition of multiple harmonics (dominantly from the first three harmonics). \\
 2) the away-side structure around $\Delta \phi=\pi$ shows an interesting double-hump shape, with a ``dip'' at $\pi$ and two ``shoulders'' around $2\pi/3$ and $4\pi/3$. This structure is mainly due to an interplay between the $v_1$ response and the $v_3$ response and also quite robust despite the two different choices of the parameter $\Sigma$. The precise shape of the away-side, though, is very sensitive to the ratio between the responses to the first and third harmonic fluctuations.\\
Qualitatively these features agree well with the experimentally observed di-hadron correlations. While certainly interesting to do, a quantitative comparison between the obtained correlation and the experimental data is not ready yet, due to a number of approximations used in the modeling and also the complication in experimental methods (e.g. observable definition, trigger and associate $p_t$ selection, the ZYAM procedure for background subtraction, etc). Some of these issues will be discussed further at the end.

One might also think about possible measurements of hard-hard correlations, i.e. di-hadron correlations with both hadrons' $p_t$ to be high enough that they are most likely from jets. Such measurements might become feasible at LHC energies when enough events with more than one pair of jets could be collected. Along similar consideration as before we may expect a component in the hard-hard azimuthal correlation due to geometric fluctuations of the form $\sim \sum_{n=1,2,3,...} 2\, (\chi^{h})^2\, < (\epsilon_n)^2>\, \cos(n\Delta\phi)$. A remnant of a near-side ridge might be visible, while the usual di-jet back-to-back correlation will be dominant.

\section{Summary}

In summary, we've studied the jet response (particularly azimuthal anisotropy) as a hard probe of the harmonic fluctuations in the initial condition of central heavy ion collisions. By implementing the fluctuations via cumulant expansion for various harmonics quantified by $\epsilon_n$ and using the geometric model for jet energy loss, we've computed the response $\chi^h_n=v_n/\epsilon_n$. Combining these results with known results for hydrodynamic response of the bulk matter expansion in the literature, we've shown that the hard-soft azimuthal correlation arising from their respective responses to the common geometric fluctuations reveals a robust and narrow near-side peak that may provide the dominant contribution to the ``hard-ridge'' observed in experimental data. 

This paper is intended to demonstrate and emphasize the main idea of hard probe for harmonic fluctuations in the initial condition, and we end by discussing a number of highly interesting aspects for further developments as well as some issues that are not fully addressed in the present paper and will be further investigated. The approach of quantifying fluctuations by cumulant expansion with one harmonic fluctuation at a time has the advantage of clearly demonstrating the response of jet energy loss to each order of harmonics, but for realistic modeling and comparison with data one needs to use Monte-Carlo generated initial matter profiles born with various harmonics and more irregularity --- this has now been implemented and the results are to be reported in a forthcoming publication. Using the Monte-Carlo initial conditions will also allow extending our study to non-central collisions as well as eliminating ambiguity due to choice of parameters. In addition the ``reaction plane'' dependence of the hard-soft correlation will be exploited with the Monte-Carlo generated fluctuations. At present the various jet energy loss models suffer from uncertainty for the energy loss in the  pre-equilibrium matter \cite{Jia:2010ee}\cite{Betz:2011tu}\cite{Blaizot:2011xf}, and it would be interesting to find ways of using geometric features of jet quenching to tightly constrain such uncertainty. While explanation of the near-side peak structure in the hard-soft di-hadron azimuthal correlation by the harmonic fluctuations appears to be robust, the full explanation of the observed away-side structure is much tricker due to known contributions from various other sources, like e.g. transverse expansion dynamics\cite{Betz:2010qh} and background effects like transverse momentum conservation and cluster correlations \cite{Bzdak:2010fd}, which all require further scrutiny. All that said, we emphasize again that the hard probe of geometric fluctuations provides a new and useful tool for exploring both the initial condition and the jet energy loss mechanism in heavy ion collisions.

\section*{Acknowledgements}
The authors thank Miklos Gyulassy, Ulrich Heinz, Jiangyong Jia, Roy Lacey, Larry McLerran, Fuqiang Wang, and Nu Xu for helpful discussions and communications. J.L. is grateful to the RIKEN BNL Research Center for partial support. The research of  X.Z. is supported under DOE Contract No. DE--FG02--87ER40365.

\end{document}